# Black Phosphorus based One-dimensional Photonic Crystals and Microcavities

Ilka Kriegel[1], Stefano Toffanin[2], Francesco Scotognella[3,4]*
[1]Department of Nanochemistry, Istituto Italiano di Tecnologia (IIT), via Morego, 30, 16163 Genova, Italy
[2]Istituto per lo Studio dei Materiali Nanostrutturati, Consiglio Nazionale delle Ricerche (CNR-ISMN), via Gobetti 101, 40129 Bologna, Italy
[3]Dipartimento di Fisica, Istituto di Fotonica e Nanotecnologie CNR, Politecnico di Milano, Piazza Leonardo da Vinci 32, 20133 Milano, Italy
[4]Center for Nano Science and Technology@PoliMi, Istituto Italiano di Tecnologia, Via Giovanni Pascoli, 70/3, 20133, Milan, Italy
* Corresponding author at: Dipartimento di Fisica, Politecnico di Milano, Piazza Leonardo da Vinci 32, 20133 Milano, Italy. E-mail address: francesco.scotognella@polimi.it

**Abstract**
The latest achievements in the fabrication of black phosphorus thin layers, towards the technological breakthrough of a phosphorene atomically thin layer, are paving the way for a their employment in electronics, optics, and optoelectronics. In this work, we have simulated the optical properties of one-dimensional photonic structures, i.e. photonic crystals and microcavities, in which few-layer black phosphorus is one of the components. The insertion of the 5 nm black phosphorous layers leads to a photonic band gap in the photonic crystals and a cavity mode in the microcavity interesting for light manipulation and emission enhancement.

**Introduction**
In the last years a lot of effort has been spent by many scientists in the materials science and solid state physics research community to obtain novel layered semiconducting materials, to tailor their properties and to integrate them as building blocks into electronic and optoelectronic devices [1,2]. Black phosphorus is particularly interesting since, in terms of electronic band gap energy, it fills the gap between graphene and metallic dichalcogenides on one side (with gaps from zero to about 0.3 eV) and semiconducting dichalcogenides on the other side (with gaps from 1 eV to 2 eV) [3–5]. In fact, black phosphorus has a band gap of 0.3 eV in the bulk and it is predicted a layer-dependent band gap [6] of up to 2 eV [7–10].
The optical properties of layered black phosphorus are characterized by strong photoluminescence [6,11] and the in-plane anisotropy [12]. In Ref. [12] the authors have measured such in-plane anisotropy by polarized optical microscopy and determined the complex refractive index dispersion in the visible range for the in-plane directions.
With the determination of the complex refractive index dispersion it is possible to design optical devices such as photonic crystals and microcavities, useful for the fabrication of optical filters, electro-optic switches and emitters with a tailored luminescence profile [13–16]. Layered semiconducting materials can be straightforwardly integrated in multilayer structures as one-dimensional (1D) photonic crystals and microcavities [17,18].
In this work, we propose the design of BP based 1D photonic crystals and microcavities. We have considered the in-plane anisotropy measured in Ref. [12] and have simulated with the transfer matrix method the transmission spectrum of the photonic structures. In the 1D photonic crystals, BP is alternated with indium tin oxide, while in the microcavity a trilayer silicon dioxide ($SiO_2$)/BP/$SiO_2$ a defect layer between two $SiO_2$/germanium dioxide ($GeO_2$) photonic crystals (usually called distributed Bragg reflectors, DBR). The arising of a photonic band gap in the photonic crystals and a cavity mode in the microcavity envisages the fabrication of BP based photonic structures for sensing and lighting applications.

**Methods**
We have used the refractive index dispersion of black phosphorus as reported in Ref. [12]. As well described in the reference, we have to consider the in-plane anisotropy, thus a refractive index in the AC direction and in the ZZ direction. In this work we refer to the black phosphorus in the AC direction with acBP, and to the black phosphorus in the ZZ direction with zzBP.
We have used the refractive index dispersion of indium tin oxide as reported in Ref. [19], the refractive index dispersion of silicon dioxide as reported in Ref. [20], and the refractive index dispersion of germanium dioxide as reported in Ref. [21] to simulate the dielectric properties of the respective layers.

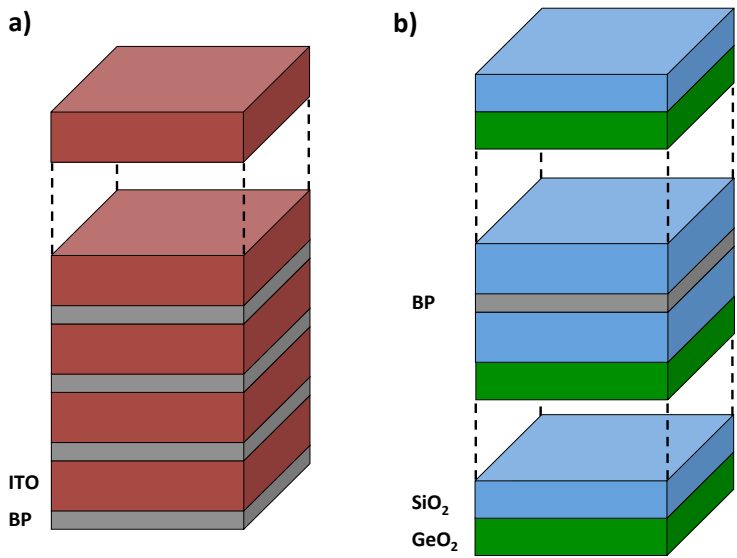

**Figure 1.** a) sketch of the BP/ITO 1D photonic crystal; b) sketch of the $(GeO_2/SiO_2)_{14.5}SiO_2/BP/SiO_2(GeO_2/SiO_2)_{14}$ microcavity.

We have considered a 1D photonic crystal in which acBP or zzBP are alternated with ITO, and a microcavity with a BP defect. The sketches of the 1D photonic crystals and the microcavity are reported in Figure 1. In the photonic crystal the thickness of black phosphorus is 5 nm as given in reference [12], while the thickness of indium tin oxide is 154 nm. The number of ac(zz)BP/ITO bilayers is 27. In the microcavity, the 1D structure is composed by a DBR of 14.5 bilayers of germanium dioxide (thickness of 90 nm) and silicon dioxide (thickness of 90 nm), a 5 nm thick layer sandwiched between two 80 nm thick layers of silicon dioxide, and a DBR of 14 bilayer of germanium dioxide and silicon dioxide. The trilayer $SiO_2/acBP/SiO_2$ acts as defect in the microcavity, as in the case of $SiO_2/MoS_2/SiO_2$ in the work reported in Ref. [17]. For the microcavity we have employed only the AC axis of BP. However, the optical properties of a microcavity embedding zzBP can be simulated in a similar way.
The transmission spectra of the photonic crystals and the microcavity have been simulated with the transfer matrix method [22–25]. We have simulated the transmission spectra in the visible range, in which the complex refractive index of BP has been determined [12].

## Results and Discussion

In Figure 2a we show the absorption spectrum of the 27 bilayer acBP/ITO photonic crystal, following the architecture depicted in Figure 1a. We have simulated the transmission spectrum and converted it to absorption ($A = -log_{10}T$). The red dashed curve indicates the absorption of the acBP/ITO photonic crystal taking into account only the real part of the refractive index of acBP. In this way, the occurrence of the photonic band gap at about 600 nm is clear. The black solid curve relates to the absorption of the same photonic crystal, but in this case the complex refractive index of acBP is taken into account, stressing the fact that in the visible range the imaginary part of the refractive index plays a significant role due to the absorption properties of BP in this wavelength range.

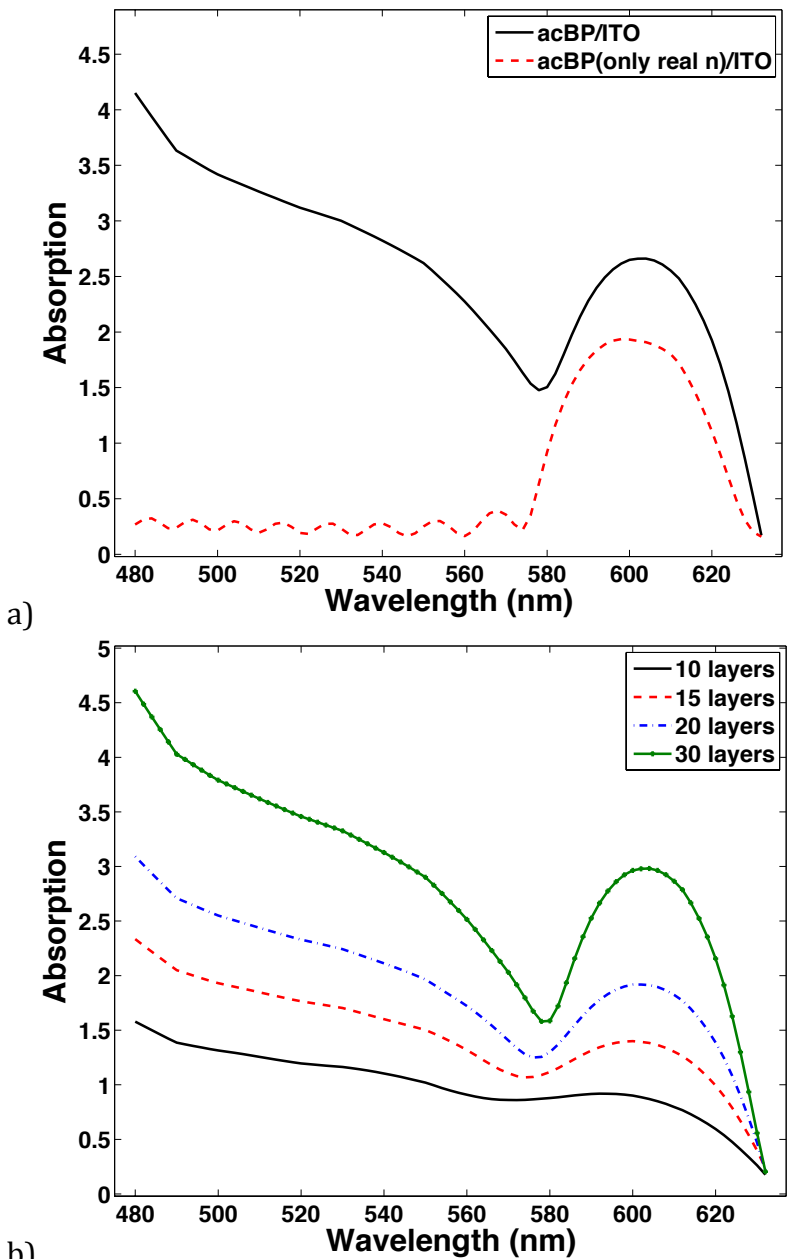

**Figure 2.** a) Transmission spectrum of 27 bilayer acBP/ITO 1D photonic crystal, where the black solid curve corresponds to the complex refractive index of acBP and the red dashed curve relates only to the real part of the refractive index of acBP; b) transmission spectra of acBP/ITO photonic crystals with different numbers of bilayers.

In Figure 2b we show the absorption spectra of the acBP/ITO photonic crystals for different numbers of acBP/ITO bilayers. The higher the number of alternating layers, the more pronounced the photonic bandgap around 600 nm.

We would like to stress that the alternation of acBP(zzBP) with ITO is very promising for a conductance of charge in the normal direction, due to good charge mobility of both materials.

Figure 3a and Figure 3b display the absorption spectra for the zzPB, highlighting that the main difference between the two acBP-based and the zzBP-based photonic crystals is in the absorption in the blue region of the spectra.

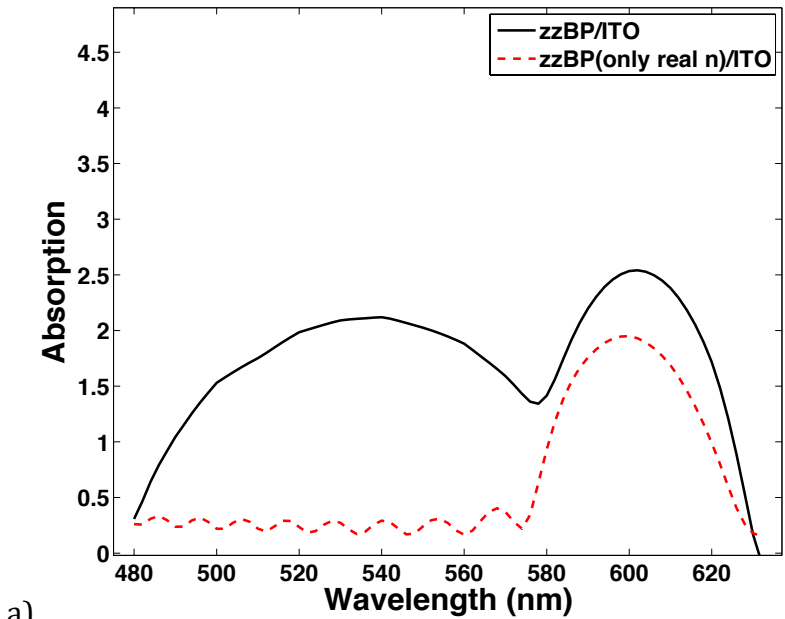
a)

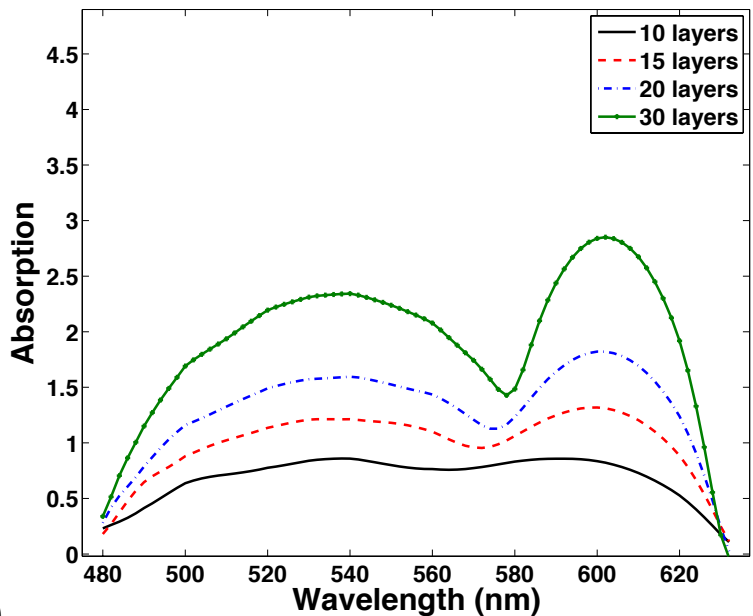
b)

**Figure 3.** a) transmission spectrum of 27 bilayer zzBP/ITO 1D photonic crystal, where the black solid curve corresponds to the complex refractive index of zzBP and the red dashed curve relates only to the real part of the refractive index of zzBP; b) transmission spectra of zzBP/ITO photonic crystals with different numbers of bilayers.

With the $(GeO_2/SiO_2)_{14.5}SiO_2/acBP/SiO_2(GeO_2/SiO_2)_{14}$ microcavity it is possible to obtain a narrow cavity mode (Figure 4). If we take into account only the real part of the refractive index of acBP, we observe and high transmissive cavity in the middle of the photonic band gap. Since the BP layer, as experimentally reported in literature [12], is very thin (5 nm), we have embedded the BP layer between two layer of $SiO_2$. This allows us to have an optical length of the defect layer, i.e. the $SiO_2$/acBP/$SiO_2$ layer, comparable to $\lambda/2$.

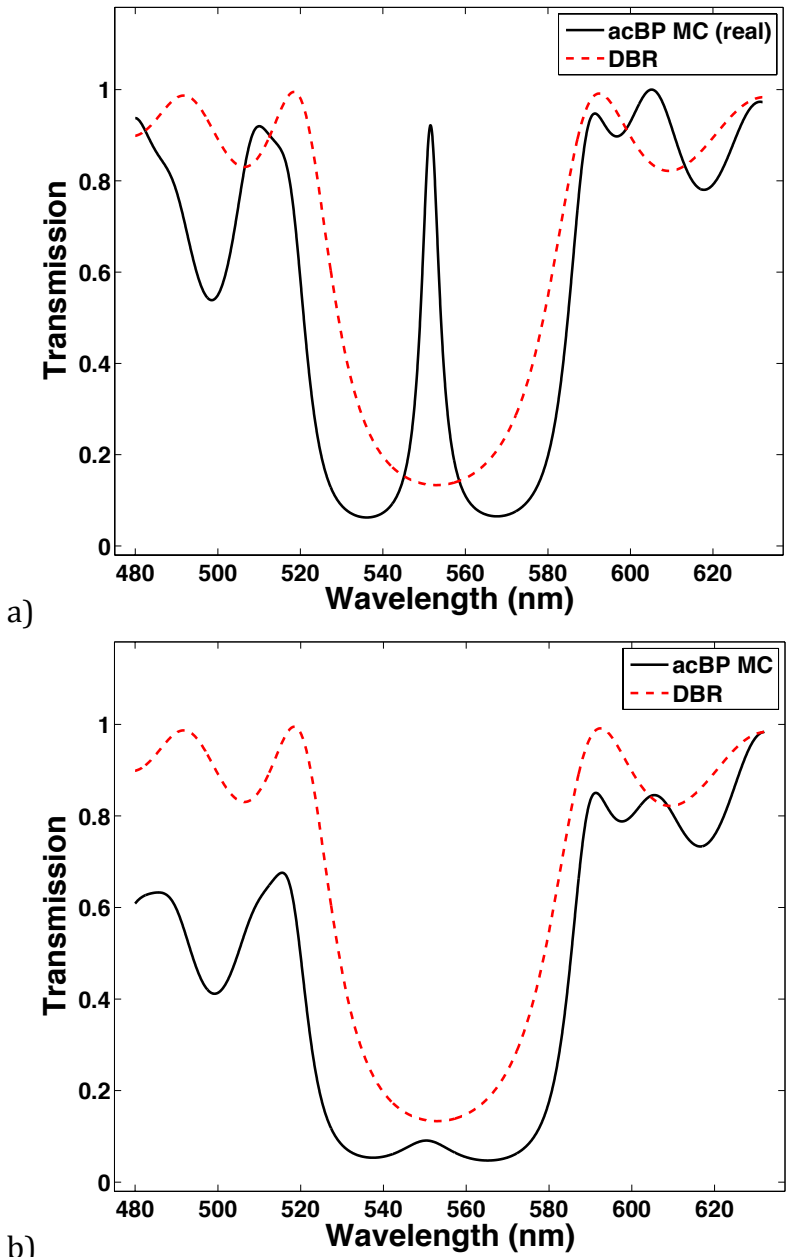

a)

b)

**Figure 4.** Transmission spectra of: a) a $(GeO_2/SiO_2)_{14.5}SiO_2/acBP/SiO_2(GeO_2/SiO_2)_{14}$ microcavity where only the real part of the refractive index is taken into account (black solid curve) and the $(GeO_2/SiO_2)_{14}$ DBR (red dashed curve); b) the same microcavity with the complex refractive index of BP (black solid curve) and the $(GeO_2/SiO_2)_{14}$ DBR (red dashed curve).

In Figure 4b the transmission spectrum takes into account the complex refractive index of acBP. The low transmission in correspondence of the cavity is due to the strong acBP absorption (i.e. imaginary part of the refractive index).
The microcavity could be very interesting for the filtering, and possibly the enhancement, of the luminescence of BP [6,11]. However, to design a proper microcavity for this purpose, the refractive index dispersion of BP in the photoluminescence range is needed, e.g. in the range between 900 nm to 1600 nm [6].

We want to stress that the realization of the multilayer BP-based photonic systems we have described in the present article is feasible and compatible with the available fabrication protocols usually implemented for exfoliating BP. In particular, oxides other than $SiO_2$ are demonstrated to form van der Waals junctions at the interface with BP. Consequently, heterostructures comprised by BP and oxides are used in the active region in planar and vertical optoelectronic devices as semiconductor only or insulator/semiconductor systems [26,27]. Typically, the BP flakes are mechanically exfoliated by using scotch tape from single crystal bulk BP and then transferred onto the substrates of interest by means of polydimethylsiloxane (PDMS) elastomer. The possibility to insert BP flakes within a planar microcavity is compatible with these fabrication protocols, as it is demonstrated by the realization of devices with oxide layers deposited on top of BP flakes (i.e. dual-gate or encapsulated top-gate field-effect transistor devices) [28,29]. Moreover, the typical deposition techniques used for fabricating multilayer PhCs such as Pulse Laser Deposition and Atomic Layer Deposition enable the deposition of high density oxides on top of soft material such as organic and inorganic thin-films [30].

Finally, the realization of the BP-based photonic systems is also compatible with the deposition of BP by means of liquid exfoliation assisted by sonication [31]. This method is expected to be highly recommended in the case of photonic and optoelectronic applications given that uniform and large size of BP sheets can be obtained and easily deposited [32]: BP active layers with lateral dimensions of the order of tens of micrometers will avoid the use of localized microscopic optical probes to characterize the performance of BP-based PhC and microcavities.

**Conclusion**

We have suggested in this work the design of acBP and zzBP based 1D photonic crystals and microcavities, by simulations based on the transfer matrix method. Taking into account the findings reported in Ref. [12] we have considered the in-plane anisotropy of BP. In the 1D photonic crystals, BP is alternated with indium tin oxide layers, ensuring a charge transport along the refractive index modulation (i.e. normal to the layer planes). Instead, in the microcavity a trilayer $SiO_2$/BP/$SiO_2$ a defect layer between two $SiO_2$/$GeO_2$ photonic crystals (usually called distributed Bragg reflectors, DBR) has been employed. A photonic band gap arising in the photonic crystals and a cavity mode in the microcavity are very promising for the fabrication of BP based photonic structures for applications as sensing and light management.